\documentstyle[emulateapj]{article} 
\received{}
\accepted{}
\journalid{ID}{DATE}
\articleid{ID}{ID}

\lefthead{Z. G. Dai and T. Lu}
\righthead{Afterglow of GRB 990123}

\begin{document}

\title{The Afterglow of GRB 990123 and a Dense Medium}

\author{Z. G. Dai\altaffilmark{1} and T. Lu\altaffilmark{2,3}}
\affil{Department of Astronomy, Nanjing University,
       Nanjing 210093, P. R. China \\
             daizigao@public1.ptt.js.cn; tlu@nju.edu.cn} 
\authoraddr{Department of Astronomy, Nanjing University,
            Nanjing 210093, P. R. China; 
	    E-mail: tlu@nju.edu.cn } 

 \altaffiltext{1} {also at Chinese Academy of Sciences - Peking University
Joint
                   Beijing Astrophysical Center, Beijing 100871, P.R. China}

\altaffiltext{2}{CCAST (World Laboratory), P.O. Box 8730, Beijing 100080,
		     P. R. China}

\altaffiltext{3}{LCRHEA, IHEP, CAS, Beijing 100039, P. R. China }

\begin{abstract}

Recent observations show that the temporal decay of the R-band afterglow
from GRB 990123 steepened about 2.5 days after the burst. We here
propose a possible explanation for such a steepening: a shock expanding
in a dense medium has undergone the transition from a relativistic phase
to a nonrelativistic phase. We find that this model is consistent
with the observations if the medium density is about $3\times 10^6\,
{\rm cm}^{-3}$. By fitting our model to the observed optical and X-ray
afterglow quantitatively, we further infer the electron and magnetic energy
fractions of the shocked medium and find that the two parameters are about
$0.1$ and $2\times 10^{-8}$ respectively. The former parameter is near
the equipartition value while the latter is about six orders of magnitude
smaller than inferred from the GRB 970508 afterglow. We also discuss
possibilities that the dense medium can be produced.

\end{abstract}

\keywords{gamma rays: bursts -- shock waves} 

\section{Introduction}

The gamma-ray burst (GRB) 990123 was an extraordinary event. It was
the brightest burst yet detected with the Wide Field Camera on the
BeppoSAX satellite (Feroci et al. 1999), and had a total gamma-ray
fluence of $\sim 5\times 10^{-4}\,{\rm erg}\,{\rm cm}^{-2}$, which
is in the top 0.3\% of all bursts. It was the first burst to be
simultaneous detected in the optical band. Optical emission with
peak magnitude of $V\sim 9$ was discovered by the Robotic Optical
Transient Search Experiment (ROTSE) during the burst and was found
to have rapidly faded down immediately after the gamma-ray emission
(Akerlof et al. 1999). The detection of the redshift showed that
the burst appears at $z\ge 1.6$ (Andersen et al. 1999; Kulkarni et al.
1999a). This implies that if the GRB emission was directed
isotropically, the inferred energy release is $\ge 1.6\times 10^{54}\,
{\rm ergs}$ (Kulkarni et al. 1999a; Briggs et al. 1999).

The burst's afterglow was detected and monitored at X-ray, optical
and radio bands. It was the brightest of all GRB X-ray afterglows
observed until now. The BeppoSAX detected the flux of the afterglow
at 2-10 keV six hours after the gamma-ray trigger to be $1.1\times
10^{-11}\,{\rm erg}\,{\rm cm}^{-2}\,{\rm s}^{-1}$ and the subsequent
temporal decay index to be $\alpha_X=-1.44\pm 0.07$ (Heise et al.
1999a, b). The R-band optical afterglow about 3.5 hours after the burst
showed a power-law decay with index $\alpha_{1R}=-1.1\pm 0.03$ (Kulkarni
et al. 1999a; Castro-Tirado et al. 1999; Fruchter et al. 1999).
This law continued until about $2.04\pm 0.46$ days after the burst.
Then the optical emission began to decline based on another power law
with index $\alpha_{2R}=-1.65\pm 0.06$ (Kulkarni et al. 1999a) or
$-1.75\pm 0.11$ (Castro-Tirado et al. 1999) or $-1.8$
(Fruchter et al. 1999). In addition, a radio flare was also
detected about 1 day after the burst (Kulkarni et al. 1999b;
Galama et al. 1999).

A scenario has been proposed to explain these observations. If the burst
is assumed to be produced from a jet, the steepening of the late optical
afterglow decay is due to the possibility that this jet has undergone the
transition from a spherical-like phase to a sideways-expansion phase
(Rhoads 1997, 1999; Kulkarni et al. 1999a; Fruchter et al. 1999; Sari,
Piran \& Halpern 1999) or that we have observed the edge of the jet
(Panaitescu \& M\'esz\'aros 1998; M\'esz\'aros \& Rees 1999).

In this {\em Letter} we propose another possible scenario, in which the
steepening of the late optical afterglow decay is due to the shock which
has evolved from a relativistic phase to a nonrelativistic phase in a dense
medium. According to the standard afterglow shock model (for a review see
Piran 1998), the afterglow is produced by synchrotron radiation or
inverse Compton scattering in the external forward wave (blast wave)
of the GRB fireball expanding in a homogeneous medium. The external
reverse shock of the fireball may lead to a prompt optical flash
(Sari \& Piran 1999). As more and more ambient matter is swept up,
the forward shock gradually decelerates and eventually
enters a nonrelativistic phase. In the meantime, the emission from
such a shock fades down, dominating at the beginning in X-rays and
progressively at optical to radio energy band. There are two limiting
cases (adiabatic and highly radiative) for the hydrodynamical evolution of
the shock. These cases have been well studied both analytically
(e.g., M\'esz\'aros \& Rees 1997; Wijers, Rees \& M\'esz\'aros 1997;
Waxman 1997a, b; Reichart 1997; Sari 1997; Vietri 1997; Katz \& Piran 1997;
M\'esz\'aros, Rees \& Wijers 1998; Dai \& Lu 1998a;
Sari, Piran \& Narayan 1998; etc) and numerically (e.g., Panaitescu,
M\'esz\'aros \& Rees 1998; Huang et al. 1998; Huang, Dai \& Lu 1998).
A partially radiative (intermediate) case has been investigated
(Chiang \& Dermer 1998; Cohen, Piran \& Sari 1998;
Dai, Huang \& Lu 1999). Here we only consider
the limiting cases. In the highly radiative model, since all
shock-heated electrons cool faster than the age of the shock, the optical
afterglow should have the same temporal decay index as the X-ray afterglow
(Sari et al. 1998), incompatible with the observations (Kulkarni
et al. 1999a). In the adiabatic model, however, the difference in the decay
index between optical and X-ray afterglows is found to be likely $1/4$,
which is consistent with the observational result $\Delta\alpha =\alpha_{1R}-
\alpha_X\approx 0.3$. This implies that the shock producing the afterglow
of GRB 990123 has evolved adiabatically. This is the starting point of our
analysis. For an adiabatic shock, the time at which it enters
a nonrelativistic phase $\propto n^{-1/3}$, where $n$ is the baryon number
density of the medium. Therefore, this time for a shock expanding in
a dense medium with density of $n\sim 10^6\,{\rm cm}^{-3}$ is two orders of
magnitude smaller than that for a shock with the same energy in a thin
medium with density of $n\sim 1\,{\rm cm}^{-3}$. Furthermore, as given in
Section 2, the afterglow at the nonrelativistic phase decays faster than
at the relativistic phase. It is natural to expect that this effect
can provide an explanation for the steepening feature of the afterglow
from GRB 990123.

Dense media have been discussed in the context of GRBs. First, Katz (1994)
suggested collisions of relativistic nucleons with a dense cloud
as an explanation of the delayed hard photons from GRB 940217. Second,
to explain the radio flare of GRB 990123, Shi \& Gyuk (1999) speculated
that a relativistic shock may have ploughed into a dense medium off the
line of sight. Third, Piro et al. (1999) and Yoshida et al. (1999)
have reported an iron emission line in the X-ray afterglow spectrum of
GRB 970508 and GRB 970828 respectively. The observed line intensity
requires a dense medium with a large iron mass concentrated in
the vicinity of the burst (Lazzati, Campana \& Ghisellini 1999).
Finally, dense media (e.g., clouds or ejecta) may appear in
the context of some energy source models, e.g., failed supernovae
(Woosley 1993), hypernovae (Paczy\'nski 1998), supranovae (Vietri \&
Stella 1998), phase transition of neutron stars to strange stars (Dai \& 
Lu 1998b), baryon decay of neutron stars (Pen \& Loeb 1998), etc.

\section{The Evolution of a Shock in a Dense Medium}

\subsection{Relativistic Phase}

Now we consider an adiabatic relativistic shock expanding in a dense
medium. The Blandford-McKee (1976) self similar solution gives the
Lorentz factor of the shock,
\begin{eqnarray}
\gamma & = & \frac{1}{4}\left[ \frac{17E(1+z)^3}{\pi nm_pc^5t_\oplus^3}
                  \right]^{1/8} \nonumber \\
      & = & 2E_{54}^{1/8}n_5^{-1/8}t_{\rm day}^{-3/8}[(1+z)/2.6]^{3/8},
\end{eqnarray}
where $E=E_{54}\times 10^{54}{\rm ergs}$ is the total isotropic energy,
$n_{5}=n/10^5\,{\rm cm}^{-3}$, $t_\oplus=t_{\rm day}\times 1\,{\rm day}$
is the observer's time since the gamma-ray trigger, $z$ is the the redshift
of the source generating this shock, and $m_p$ is the proton mass.

In analyzing the spectrum and light curve of synchrotron radiation
from the shock, one needs to know two crucial frequencies:
the synchrotron radiation peak frequency ($\nu_m$) and the cooling
frequency ($\nu_c$). In the standard afterglow shock picture, the electrons
heated by the shock are assumed to have a power-law distribution:
$dN_e/d\gamma_e\propto \gamma_e^{-p}$ for $\gamma_e\ge\gamma_{em}$,
where $\gamma_e$ is the electron Lorentz factor and the minimum Lorentz
factor $\gamma_{em}=610\epsilon_e\gamma$. The power-law index
$p\approx 2.56$ by fitting the spectrum and light curve of the observed
afterglow of GRB 990123 (see below). We further assume that $\epsilon_e$
and $\epsilon_B$ are ratios of the electron and magnetic energy
densities to the thermal energy density of the shocked medium respectively.
Based on these assumptions, the synchrotron radiation peak frequency
in the observer's frame can be written as
\begin{eqnarray}
\nu_m & = & \frac{\gamma\gamma_{em}^2}{1+z}\frac{eB'}{2\pi m_ec} \nonumber \\
     & = & 8.0\times 10^{11}\epsilon_e^2\epsilon_{B,-6}^{1/2}
               E_{54}^{1/2}t_{\rm day}^{-3/2} \nonumber \\ 
          &  & \times [(1+z)/2.6]^{1/2}\,\,{\rm Hz},
\end{eqnarray}
where $\epsilon_{B,-6}=\epsilon_B/10^{-6}$ and $B'=(32\pi \epsilon_B\gamma^2
nm_pc^2)^{1/2}$ is the internal magnetic field strength of the shocked
medium. According to Sari et al. (1998), the cooling
frequency, the frequency of electrons with Lorentz factor of $\gamma_c$
that cool on the dynamical time of the shock, is given by
\begin{eqnarray}
\nu_c & = & \frac{\gamma\gamma_c^2}{1+z}\frac{eB'}{2\pi m_ec}
    =   \frac{18\pi em_ec(1+z)}{\sigma_T^2B'^3\gamma t_\oplus^2} \nonumber \\
      & = & 1.9\times 10^{16}\epsilon_{B,-6}^{-3/2}E_{54}^{-1/2}
        n_5^{-1}t_{\rm day}^{-1/2}\nonumber \\
           &  &   \times [(1+z)/2.6]^{-1/2}\,\,{\rm Hz},
\end{eqnarray}
where $\sigma_T$ is the Thompson scattering cross section. From equations (2)
and (3), Sari et al. (1998) have further defined two critical times,
when the breaking frequencies $\nu_m$ and $\nu_c$ cross the observed
frequency $\nu=\nu_{15}\times 10^{15}\,{\rm Hz}$:
$t_m=8.6\times 10^{-3}\epsilon_e^{4/3}\epsilon_{B,-6}^{1/3}
    E_{54}^{1/3}[(1+z)/2.6]^{1/3}\nu_{15}^{-2/3}\,{\rm days}$,
and $t_c=380\epsilon_{B,-6}^{-3}E_{54}^{-1}
     n_5^{-2}[(1+z)/2.6]^{-1}\nu_{15}^{-2}\,{\rm days}$.
Therefore we see that for $E_{54}\sim 1.6$,
$\epsilon_e\sim 0.1$, $\epsilon_{B,-6}\sim 0.02$, and $n_5\sim 30$
inferred in the next section, the optical afterglow in several days
after the burst should result from those slowly-cooling
electrons and the X-ray afterglow from those fastly-cooling electrons.

The observed synchrotron radiation peak flux can be obtained by
\begin{eqnarray}
F_{\nu_m} & = & \frac{N_e\gamma P'_{\nu_m}(1+z)}{4\pi D_L^2} \nonumber \\
         & = & 4.2\epsilon_{B,-6}^{1/2}E_{54}n_5^{1/2}[(1+z)/2.6]
          D_{L,28}^{-2}\,\,{\rm Jy},
\end{eqnarray}
where $N_e$ is the total number of swept-up electrons, $P'_{\nu_m}
=m_ec^2\sigma_TB'/(3e)$ is the radiated power per electron per unit
frequency in the frame comoving with the shocked medium, and
$D_L=D_{L,28}\times 10^{28}\,{\rm cm}$ is the distance to the source.
In the light of equations (2)-(4), one can easily find
the spectrum and light curve of the afterglow,
\begin{equation}
F_\nu=\left \{
       \begin{array}{llll}
         (\nu/\nu_m)^{-(p-1)/2}F_{\nu_m}\\ \,\,\,\,\,\,\,\, 
                         \propto \nu^{-(p-1)/2}
              t_\oplus^{3(1-p)/4}\,\,\,\,\,\, {\rm if}\,\, \nu_m<\nu<\nu_c; \\
         (\nu_c/\nu_m)^{-(p-1)/2}(\nu/\nu_c)^{-p/2}F_{\nu_m}
                 \\ \,\,\,\,\,\,\,\, \propto\nu^{-p/2}t_\oplus^{(2-3p)/4}
                     \,\,\,\,\,\,\,\,\,\,\,\,\,\,\,\, {\rm if}\,\, \nu>\nu_c,
        \end{array}
       \right.
\end{equation}
where the low-frequency radiation component has not been considered
(Sari et al. 1998). In the GRB 990123 case,
we require $\nu_m<\nu<\nu_c$ for the optical afterglow and
$\nu>\nu_c$ for the X-ray afterglow. Thus, the R-band afterglow
decay index $\alpha_R=3(1-p)/4$ and the X-ray decay index $\alpha_X=
(2-3p)/4$, which are well consistent with the observational results
$\alpha_{1R}=1.1\pm 0.03$ and $\alpha_X=-1.44\pm 0.07$ if $p\approx 2.56$.

\subsection{Nonrelativistic Phase}

As it sweeps up sufficient ambient matter, the shock will eventually go
into a nonrelativistic phase. During such a phase, the shock's velocity
$v\propto t_\oplus^{-3/5}$, its radius $r\propto t_\oplus^{2/5}$,
the internal field strength $B'\propto t_\oplus^{-3/5}$ and the typical
electron Lorentz factor $\gamma_{em}\propto t_\oplus^{-6/5}$. Thus,
we obtain the synchrotron peak frequency $\nu_m\propto \gamma_{em}^2B'
\propto t_\oplus^{-3}$, the cooling frequency $\nu_c\propto B'^{-3}
t_\oplus^{-2}\propto t_\oplus^{-1/5}$ and the peak flux $F_{\nu_m}\propto
N_eP'_{\nu_m}\propto r^3B'\propto t_\oplus^{3/5}$. According to these
scaling laws, we further derive the spectrum and light curve at the
nonrelativistic stage:
\begin{equation}
F_\nu=\left \{
       \begin{array}{llll}
         (\nu/\nu_m)^{-(p-1)/2}F_{\nu_m} \\ \,\,\,\,\, \propto \nu^{-(p-1)/2}
              t_\oplus^{(21-15p)/10}\,\,\,\, {\rm if}\,\, \nu_m<\nu<\nu_c; \\
         (\nu_c/\nu_m)^{-(p-1)/2}(\nu/\nu_c)^{-p/2}F_{\nu_m}
                 \\ \,\,\,\,\, \propto\nu^{-p/2}t_\oplus^{(4-3p)/2}
            \,\,\,\,\,\,\,\,\,\,\,\,\,\,\,\,\,\,\,\,\, {\rm if}\,\, \nu>\nu_c.
        \end{array}
       \right.
\end{equation}
From equation (6), we can see the R-band decay index $\alpha_R=(21-15p)/10$
for radiation from slowly-cooling electrons or $\alpha_R=(4-3p)/2$ for
radiation from rapidly-cooling electrons. If $p\approx 2.56$, then $\alpha_R
\approx -1.74$ or $-1.84$, in excellent agreement with the observations
in the time interval of 2.5 days to 20 days after the burst
(Kulkarni et al. 1999a; Fruchter et al. 1999; Castro-Tirado et al. 1999).

\section{Constraints on Parameters}

In the above section, we show that as an adiabatic shock expands
in a dense medium from an ultrarelativistic phase to a nonrelativistic
phase, the decay of the radiation from such a shock will steepen. This
effect may fit the observed steepening better than the alternative
interpretation --- jet sideways expansion. In the latter interpretation,
the temporal decay of a late afterglow is very likely to be $\propto
t_\oplus^{-p}$ (Rhoads 1997, 1999; Sari et al. 1999). We further
analyze our effect and infer some parameters of the model.

According to the analysis on the R-band light curve of the GRB 990123
afterglow (Kulkarni et al. 1999a; Fruchter et al. 1999;
Castro-Tirado et al. 1999), the observed break occurred at
$t_\oplus = 2.04\pm 0.46\,$days. This implies $\gamma\sim 1$
at $t_{\rm day}\approx 2.5$. From equation (1), therefore, we find
$n_5\sim 16E_{54}$, where the redshift $z=1.6$ has been used. We now
continue to consider two observational results. First, on January
23.577 UT, the Palomar 60-inch telescope detected the R-band magnitude
$R=18.65\pm 0.04$, corresponding to the flux $F_R\sim
100\,\mu{\rm Jy}$ at $t_{\rm day}\approx 0.17$ (Kulkarni et al. 1999a).
Considering this result in equation (5) together with equations
(2) and (4), we can derive
\begin{equation}
\epsilon_e^{p-1}\epsilon_{B,-6}^{(p+1)/4}E_{54}^{(p+3)/4}
n_5^{1/2} \sim 0.01,
\end{equation}
where the right number has been obtained by taking $p\approx 2.56$
and $D_{L,28}\sim 3.7$. Second, on January 24.65 UT,
the BeppoSAX observed the X-ray (2-10\,keV)
flux $F_X\sim 5\times 10^{-2}\,\mu$Jy (Heise et al. 1999a, b). Combining
this result with equations (2)-(5), we can also derive
\begin{equation}
\epsilon_e^{p-1}\epsilon_{B,-6}^{(p-2)/4}E_{54}^{(p+2)/4}\sim 0.03.
\end{equation}
Since $E_{54}\sim 1.6$ (Briggs et al. 1999; Kulkarni et al. 1999a),
the medium density $n_5 \sim 30$ and the solution of equations (7) 
and (8) is $\epsilon_e\sim 0.1$ and $\epsilon_{B,-6}\sim 0.02$.
Our inferred value of $\epsilon_e$ is near the equipartition value,
in agreement with the result of Wijers \& Galama (1998) and Granot,
Piran \& Sari (1998), while our $\epsilon_B$ is about six orders of
magnitude smaller than the value inferred from the afterglow
of GRB 970508. Of course, the field density for GRB 971214
has been estimated to be less than $10^{-5}$ times the equipartition
value (Wijers \& Galama 1998). As suggested by Galama et al. (1999),
such differences in field strength may reflect differences in energy
flow from the central engine.

\section{Discussion and Conclusion}

In the above section, we find the medium density $n\sim 3\times
10^6\,{\rm cm}^{-3}$ for our model to fit the observed optical and
X-ray afterglow of GRB 990123. Now we show that even in the the presence
of such a dense medium, the optical and X-ray radiations from the forward
shock were neither self absorbed in the shocked medium nor scattered
in the unshocked medium. First,  the self-absorption frequency of the
shocked medium is (Wijers \& Galama 1998; Granot et al. 1998)
$\nu_a\sim 10^3\,{\rm GHz}(\epsilon_e/0.1)^{-1}(\epsilon_{B,-6}
/0.01)^{1/5}E_{54}^{1/5}(n_5/10)^{3/5}$. This estimate should be
the upper limit because of the presence of a possible low-energy
electron population (Waxman 1997b). Clearly, $\nu_a$ is much less
than the optical frequency, implying that the self absorption in
the shocked medium didn't affect the optical and X-ray afterglow.
In fact, this estimate is valid only for $\nu_a < \nu_m$.
When $\nu_a>\nu_m$, $\nu_a$ must have decayed. As a result,
the flux at 8.46 GHz first increased as $t_\oplus^{1.25}$
and then declined as $t_\oplus^{-1.74}$ for $\nu_a<8.46$ GHz
during the nonrelativistic phase. This might provide an explanation for
the observed radio flare. Second, a photon emitted from the shock may be
scattered by the electrons in the unshocked medium. The scattering optical
depth $\tau\sim\sigma_TnR$ (where $R$ is the typical radius of the medium).
If the medium was distributed isotropically and homogeneously and its
mass $M\sim 10M_\odot$ (the typical mass of a supernova ejecta),
then $\tau\sim 0.05(M/10M_\odot)^{1/3}(n_5/10)^{2/3}\ll 1$.
This implies that the afterglow from the shock was hardly affected
by the medium.

For other well-studied afterglows, e.g., GRB 970228 and GRB 970508,
their ambient densities must be very low for three reasons: (i) In these
bursts there was no observed break in the optical light curve as long as
the afterglow could be observed (Fruchter et al. 1998; Zharikov et al. 1998).
(ii) The fluctuation appearing in the radio afterglow light curve
of GRB 970508 requires the shock had been relativistic for
several weeks (Waxman, Kulkarni \& Frail 1998). (iii) The analysis of the
afterglow spectrum of GRB 970508 leads to a low ambient density
$n<10\,{\rm cm}^{-3}$ (Wijers \& Galama 1998; Granot et al. 1998).
However, the observed iron emission line in the X-ray afterglow
spectrum of GRB 970508 indeed requires a dense medium with density
$\sim 10^9\,{\rm cm}^{-3}$ (Lazzati et al. 1999).
The only way to reconcile a monthly lasting power-law afterglow
with iron line emission is through a particular geometry, in which
the line of sight is devoid of the dense medium. In contrast to
this idea, we suggest that for GRB 990123 a dense medium of
$n\sim 3\times 10^6\,{\rm cm}^{-3}$ appears at least at the line
of sight or perhaps isotropically.

How was the dense medium produced? One possibility was a cloud and
another possibility was an ejecta from the GRB site. There have been
several source models (mentioned in Introduction) in the literature which
may lead to massive ejecta. Here we want to discuss one of them in detail.
Timmes, Woosley \& Weaver (1996) showed that Type II supernovae may
produce a kind of neutron star with $\sim 1.73M_\odot$. If these massive
neutron stars have very short periods at birth, they may subsequently
convert into strange stars due to rapid loss of angular momenta
(Cheng \& Dai 1998), and perhaps the strange stars are
differentially rotating (Dai \& Lu 1998b). Even though this model
is somewhat similar to the supranova model of Vietri \& Stella (1998),
resultant compact objects are strange stars in our model
and black holes in the supranova model. We further discuss implications
of our model. First, the model leads to low-mass loading matter
because of thin baryonic crusts of the strange stars. Second,
such stars result in GRBs with spiky light curves,
being consistent with the analytical result from the observed
data of GRB 990123 (Fenimore, Ramirez-Ruiz \& Wu 1999). The third
advantage of this model is to be able to explain well the property of
the early afterglow of GRB 970508 by considering energy injection from the
central pulsar (Dai \& Lu 1998b, c). Finally, a dense medium, the supernova
ejecta, appears naturally.

Our scenario proposed in this {\em Letter} requires a dense medium
with density $\sim 3\times 10^6\,{\rm cm}^{-3}$ to explain the steepening
in the temporal decay of the R-band afterglow about 2.5 days
after GRB 990123. We also suggest that this medium could be a 
supernova/supranova/hypernova ejecta. Thus, if the mass of 
the medium is assumed to be $M\sim 10M_\odot$,
its radius can be estimated to be $R\sim 3\times 10^{17}\,{\rm cm}
(M/10M_\odot)^{1/3}(n_5/10)^{-1/3}$. According to equation (1), we can 
integrate $dr=2\gamma^2 cdt_\oplus$ and thus find that the postburst
2.5-day time in the observer's frame corresponds to about 20 days
in the unshocked medium's frame. This implies that the radius at which
the shock entered a nonrelativistic phase is about $5\times 10^{16}$ cm.
This radius is much less than that of the medium. Therefore,
the medium discussed here was so wide and dense that the
ultrarelativistic shock must have become nonrelativistic about
2.5 days after the burst.

In summary, a simple explanation for the ``steepening''
observed in the temporal decay of the late R-band afterglow of
GRB 990123 is that a shock expanding in a dense medium with density
of $\sim 3\times 10^6\, {\rm cm}^{-3}$ has evolved from a relativistic
phase to a nonrelativistic phase. We find that this scenario not only
explains well the optical afterglow but also accounts for the observed
X-ray afterglow quantatitively.

\acknowledgments

We would like to thank J. I. Katz, S. R. Kulkarni, A. Mitra and
the anonymous referee for invaluable suggestions, and Y. F. Huang and
D. M. Wei for helpful discussions. This work was supported by the National
Natural Science Foundation of China (grants 19825109 and 19773007).

\end{document}